\documentclass[fleqn,usenatbib]{mnras}
\usepackage{newtxtext,newtxmath}
\usepackage[T1]{fontenc}
\usepackage{graphicx}	% Including figure files
\usepackage{amsmath}	% Advanced maths commands
\usepackage{amssymb}	% Extra maths symbols

\usepackage{amsmath}
\usepackage{amsfonts}
\usepackage{amssymb}
\usepackage{graphicx}
\usepackage{hyperref}
\usepackage{booktabs}
\usepackage{cleveref}
\usepackage{color}
\usepackage{changes}
\usepackage{enumerate}

\newcommand{\Msun}{\,{\rm M_\odot}}
\newcommand{\Mblack}{M_\bullet}

\begin{document}

%\title[Supermassive black holes and $H_0$]{Consequences of the $H_0 $ measurements  on growth of black holes across the cosmic time}

\title[Effects of $H_0$ on the Growth of $z > 6.5$ Quasars]{Effects of the Hubble Parameter on the Cosmic Growth of the First Quasars}

\author[R. C. Nunes, F. Pacucci]{
Rafael C. Nunes$^{1}$\thanks{E-mail: rafadcnunes@gmail.com},
Fabio Pacucci$^{2,3}$\thanks{BHI \& Clay Fellow}
\\
% List of institutions
$^{1}$Divis\~ao de Astrof\'isica, Instituto Nacional de Pesquisas Espaciais, Avenida dos Astronautas 1758, S\~ao Jos\'e dos Campos, 12227-010, SP, Brazil\\
$^{2}$Black Hole Initiative, Harvard University,
Cambridge, MA 02138, USA\\
$^{3}$Center for Astrophysics $\vert$ Harvard \& Smithsonian,
Cambridge, MA 02138, USA\\
}

% These dates will be filled out by the publisher
\date{Accepted XXX. Received YYY; in original form ZZZ}

% Enter the current year, for the copyright statements etc.
\pubyear{2020}

% Don't change these lines

\label{firstpage}
\pagerange{\pageref{firstpage}--\pageref{lastpage}}
\maketitle

\label{firstpage}
\pagerange{\pageref{firstpage}--\pageref{lastpage}}

\begin{abstract}
Supermassive black holes (SMBHs) play a crucial role in the evolution of galaxies and are currently detected up to $z\sim 7.5$. Theories describing black hole (BH) growth are challenged by how rapidly seeds with initial mass $\Mblack \lesssim 10^5 \Msun$, formed at $z \sim 20-30$, grew to $\Mblack \sim 10^9 \Msun$ by $z\sim 7$. 
Here we study the effects of the value of the Hubble parameter, $H_0$, on models describing the early growth of BHs. First, we note that the predicted mass of a quasar at $z=6$ changes by $> 300\%$ if the underlying Hubble parameter used in the model varies from $H_0 = 65$ to $H_0 = 74$ km s$^{-1}$Mpc$^{-1}$, a range encompassing current estimates. Employing an MCMC approach based on priors from $z \gtrsim 6.5$ quasars and on $H_0$, we study the interconnection between $H_0$ and the parameters describing BH growth: seed mass $M_i$ and Eddington ratio $f_{\rm Edd}$.
Assuming an Eddington ratio of $f_{\rm Edd} = 0.7$, in agreement with previous estimates, we find $H_0 = 73.6^{+1.2}_{-3.3}$ km s$^{-1}$Mpc$^{-1}$. In a second analysis, allowing all the parameters to vary freely, we find $\log(M_{i}/M_{\odot}) > 4.5$ (at 95\% CL), $H_0 = 74^{+1.5}_{-1.4}$ km s$^{-1}$Mpc$^{-1}$ and $f_{\rm Edd}=0.77^{+0.035}_{-0.026}$ at 68\% CL. Our results on the typical Eddington ratio are in agreement with previous estimates. Current values of the Hubble parameter strongly favour heavy seed formation scenarios, with $M_i \gtrsim 10^4 \Msun$. In our model, with the priors on BH masses of quasars used, light seed formation scenarios are rejected at $\sim 3\sigma$.

\end{abstract}


\section{Introduction}
\label{Introduction}

Numerous surveys of the high-redshift Universe ($z\gtrsim 6$) strongly suggest that supermassive black holes (SMBHs), with masses in the range $10^{6-10} \Msun$, are already in place by that cosmic age and provide the energy to power quasars (e.g., \citealt{Fan_2006, Wu_2015, Banados_2018}). The detection of several SMBHs at redshift $z \gtrsim 7$ with masses $\sim 10^9 \Msun$ is a significant challenge to the standard model of black hole (BH) growth: in fact, it is still unclear how did these BHs form and grow so rapidly over cosmic time. Current theories describe the first BH seeds to form at $z \sim 20-30$, less than $\sim 200$ Myr after the Big Bang, and then to rapidly grow, by gas accretion and mergers, to their final masses \citep{Pacucci_2020_acc_mer}. Extensive reviews about the formation and early growth of quasars can be found in \cite{2017PASA...34...22G}, \cite{2019ffbh.book.....L}, and \cite{2019arXiv191105791I}.

Over the past two decades, a large number of high-$z$ quasars have been discovered in surveys as SDSS and the CFHQS \citep{Jiang_2009,Willott_2010}, PanSTARRS1 \citep{2016ApJS..227...11B},
VST/ATLAS \citep{2015MNRAS.451L..16C}, DES \citep{2015MNRAS.454.3952R,2017MNRAS.468.4702R}, Subaru/HSC \citep{Matsuoka_2016}, UKIDSS \citep{2007MNRAS.376L..76V} and VIKING \citep{2015ApJ...801L..11V, 2013ApJ...779...24V}. Growing observational and theoretical evidence strongly suggest that the seeds at the origin of these massive objects formed at early times, likely at $z \sim 20-30$ \citep{BL01}. 

One possibility, the ``light seeds model'', consists in these seeds being formed as remnants of the first population of stars (i.e., Population III, or Pop III, stars). While large uncertainties remain on the initial mass function of Pop III stars, several simulations and theoretical models point to a mass of the BH remnant in the range $10 \lesssim \Mblack/M_{\odot} \lesssim 1000$ (e.g., \citealt{Hirano_2014}). Alternatively, the ``heavy seeds model'' predicts the existence of more massive BHs, with a typical mass scale $\sim 10^5 \Msun$ already at formation. These heavy seeds are named  direct  collapse  black  holes  (DCBHs;  e.g., \citealt{Bromm_Loeb_2003, Lodato_Natarajan_2006, Pacucci_2017_Seed}). 

While heavy seeds could reach the $\sim 10^9 \Msun$ mass scale in time to match the observations of $z \sim 7$ quasars with Eddington-limited accretion, light seeds most likely need episodes of super-Eddington accretion \citep{Haiman_2001,Volonteri_2005,2007ApJ...665..107P,2009ApJ...696.1798T, Madau_2014_super, Volonteri_2015, PVF_2015, Begelman_Volonteri_2017, Regan_2019}.

In addition to these two baseline formation channels, additional scenarios have been proposed, such as BHs formed from stellar collisions \citep{Devecchi_2009, Devecchi_2012, Katz_2015} and black hole mergers \citep{Davies_2011, Lupi_2014}. 

Currently, six quasars are known at $z > 7$ \citep{2019arXiv191105791I}.
The farthest one thus far is J1342+0928 at $z = 7.54$ with $\sim 7.8 \times 10^8 \Msun$ \citep{Banados_2018}. Future surveys in the electromagnetic spectrum  like Lynx \citep{2018arXiv180909642T}, AXIS \citep{AXIS}, Athena \citep{2020AN....341..224B} and the James Webb Space Telescope, as well as surveys in the gravitational wave realm, e.g. LISA \citep{2017arXiv170200786A}, will provide invaluable information about the formation and the growth process of high-$z$ BH seeds \citep{Pacucci_2020_acc_mer}. 

Remarkably, any prediction on how BHs grow over cosmic time depends on the underlying cosmology assumed. A modification in the value of the cosmological parameters used, and/or any extension beyond the concordance $\Lambda$CDM cosmology, could significantly change the evolution and the dynamics of the Universe, possibly producing very different predictions for BH growth. Particular attention should be granted to estimates of the Hubble parameter, $H_0$, which describes the current expansion rate of the universe. The most recent analyses of the Cosmic Microwave Background (CMB) observations by the Planck collaboration, assuming the $\Lambda$CDM scenario, obtained $H_0=67.4 \pm 0.5$ km s$^{-1}$Mpc$^{-1}$ \citep{2018arXiv180706209P}. A model-independent local measurement by the Hubble Space Telescope (HST) suggested instead $H_0= 74.03 \pm 1.42$ km s$^{-1}$Mpc$^{-1}$ \citep{Riess_2019}, which is in $4.4\sigma$ tension with Planck's estimate. Additionally, the H0LiCOW collaboration reports $H_0= 73.3^{+1.7}_{-1.8}$ km s$^{-1}$Mpc$^{-1}$ \citep{2019arXiv190704869W}. Another accurate independent measure was carried out in \cite{2019ApJ...882...34F}, showing that $H_0 = 69.8 \pm 0.8$ km s$^{-1}$Mpc$^{-1}$. These are the most robust estimates of $H_0$ available in literature. As noted, there is a high degree of statistical divergence between them. This observed tension could be a signal of additional fundamental new physics beyond the standard $\Lambda$CDM model (see, e.g., \citealt{2019NatAs...3..891V, 2019EPJC...79..576K} and references therein).

Here, we aim to understand the effect of the value of $H_0$ on models for early BH growth. First, we show how the predicted mass at $z=6$ can be significantly affected by the choice of $H_0$. Then, we constrain $H_0$ using information from the mass of the farthest quasars detected thus far, in the range $ 6.5 < z < 7.54$, assuming that mass growth occurs mostly by gas accretion. Conversely, we then study how much $H_0$, assumed a free parameter, can influence our estimate of the parameters that quantify the BH growth.

This study is organized as follows. In Sec. 2 we revise the theoretical model used to describe the evolution of BH mass over cosmic time. In Sec. \ref{data} we present our data sets and in Sec. \ref{results} our main results and discussions. Finally, Sec. \ref{conclusion} summarizes our conclusions and presents some future perspectives.

\section{The cosmic growth of black holes}
\label{model}

In this section we review the theoretical framework used to describe the cosmic growth of BHs via gas accretion, from seed formation ($z \sim 20-30$) to the observation of the farthest quasars ($z \gtrsim 6.5$). The formalism adopted here is described extensively in \cite{PVF_2015} and \cite{Pacucci_2017} (see also, e.g., \citealt{Shapiro_2005, Ricarte_2018}).

The evolution in cosmic time $t$ of the BH mass $\Mblack$, starting from an initial mass of the seed $M_i$, is usually described with the following set of three parameters:

\begin{enumerate}

\item The matter-to-energy conversion efficiency factor $\epsilon$, which describes the fraction of rest-mass energy that is radiated away during gas accretion. The efficiency factor $\epsilon$ is defined as:
\begin{equation}
\epsilon = \frac{L}{\dot{M} c^2} \, ,
\end{equation}
where $\dot{M}$ is the accretion rate onto the BH and $L$ is its luminosity. The factor $\epsilon$ is customarily assumed to be $\sim10\%$  for  radiatively  efficient  accretion  disks  \citep{Shakura_Sunyaev_1976}. In case of optically thick accretion disks, or radiatively inefficient accretion flows (or RIAF), the factor $\epsilon$ can be significantly lower (see, e.g., \citealt{Narayan_McClintock_2013}).

\item The Eddington ratio, which parametrizes the accretion rate $\dot{M}$ on a BH of mass $\Mblack$ in terms of the Eddington accretion rate $\dot{M}_{\rm Edd} \approx 2.2 \times 10^{-8} (\Mblack/\Msun) \, \mathrm{\Msun \, yr^{-1}}$:
\begin{equation}
f_{\rm Edd} = \frac{\dot{M}}{\dot{M}_{\rm Edd}} \, .
\end{equation}

\item The duty cycle ${\cal D}$, quantifying the fraction of time spent accreting (i.e., the continuity of mass inflow). It is worth noting that typical quasar timescales are of order $\sim 100 \, \mathrm{Myr}$.
\end{enumerate}

If we assume that the accretion rate is dominated by baryonic matter, then the BH growth rate is found with the following expression (see, e.g., \citealt{Shapiro_2005, Pacucci_2017}):

\begin{equation}
\label{M_evolution}
\dot{M} = \frac{{\cal D} f_{Edd} (1 - \epsilon)}{\epsilon} \frac{M}{\tau} \, ,
\end{equation}
where $\tau$ is the characteristic accretion timescale, or Salpeter timescale \citep{Salpeter_1955}, $\tau \approx 0.45$ Gyr.
 
In general, the matter-to-energy conversion efficiency factor $\epsilon$ is a strong function of the BH spin (e.g., \citealt{Bardeen_1970, Novikov_Thorne, Narayan_McClintock_2013}). The efficiency for disk accretion onto a Schwarzschild (i.e., non-rotating) BH is $\epsilon = 0.057$, while for a Kerr, maximally rotating BH the value is found to be $\epsilon \sim 0.32$. In fact, for rotating BHs the accretion disk extends farther inwards, closer to the event horizon, so that a larger fraction of its energy can be radiated away. As we do not track the spin evolution in our work, in what follows we assume $\epsilon = 0.1$.

If we assume that $\epsilon$, ${\cal D}$ and $f_{\rm Edd}$ are constant between $t_i$ and $t$, we can easily integrate Eq. (\ref{M_evolution}), obtaining:

\begin{equation}
\label{M}
M(t) = M(t_i) \exp \Big[ \frac{{\cal D} f_{\rm Edd}(1-\epsilon)}{\epsilon} \frac{t-t_{\rm i}}{\tau} \Big] \, ,  
\end{equation}
where $t_{\rm i}$ is the initial time when the BH has a mass $M_i$.
The lookback time as a function of $z$ can be written as
\begin{equation}
t_i(z_0) = t(z_0) - t(z_i) = \frac{1}{H_0} \int_{z_0}^{z_i} \frac{dz'}{(1+z') E(z')} \, ,
\end{equation}
where $E(z) = H(z)/H_0$ is the ratio between the Hubble parameter at $z$ and its current value. The value $t_i(z_0)$ is the age of the object at redshift $z_0$, assuming that it formed at redshift $z_i$. For our purposes, $z_0$ is the redshift where the quasar is detected, and $z_i$ is the redshift at seed formation. 

Assuming that BH growth occurred only via baryonic gas accretion is certainly an approximation. A more realistic scenario would allow for additional growth channels, e.g. black hole mergers and contributions from dark matter (collisionless and self-interacting species). Recent studies (e.g., \citealt{Pacucci_2020_acc_mer}) have shown that gas accretion is significantly dominant over mergers for the growth of BHs with $\Mblack \gtrsim 10^5 \Msun$ in the early Universe ($z \gtrsim 6$). For this reason, we believe that our model including only gas accretion is, to the first order, a good approximation of how BHs grew at early cosmic epochs.

\section{Methodology and data set}
\label{data}

In the following we briefly describe the data sets we use to explore the parameter space of our model.
\\

\textbf{Mass estimates for $z \geq 6.5$ quasars}: We consider quasars in the redshift range $z$ $\in$ [6.5, 7.54], i.e. up to the farthest quasar detected thus far. In particular, we consider the following sources: J2348-3054 ($z = 6.9018$, \citealt{2016ApJ...816...37V}), J0109-3047 ($z = 6.7909$, \citealt{2016ApJ...816...37V}), J0305-3150 ($z = 6.6145$, \citealt{2016ApJ...816...37V}),  P036+03 ($z = 6.5412$, \citealt{2015ApJ...801L..11V,2015ApJ...805L...8B}), J1342+0928 ($z=7.541$, \citealt{2018Natur.553..473B}), J1243+0100 ($z=7.07$, \citealt{Matsuoka_2019}), J1120+0641 ($z =7.085$, \citealt{2011Natur.474..616M}) and J0038-1527 ($z = 7.021$, \citealt{Wang_2018}). This  information is summarized in Table \ref{tab:SMBH_data}.
Hence, we include in our analysis all $z >7$ quasars from \cite{2019arXiv191105791I} for which an error bar for the mass is reported, and some quasars in the range $z$ $\in$ [6.5, 7.0] for which the error bars are small, i.e. the mass is known typically within a factor $\sim 2$. There are currently $\sim 16$ confirmed quasars within $z$ $\in$ [6.5, 7.0] (see the NASA/IPAC Extragalactic Database, NED, and, e.g., \citealt{Matsuoka_2016}) but some of these sources are characterized by very large uncertainties for the BH mass, or it is unconstrained altogether. In order to maximize the accuracy of our Markov Chain Monte Carlo method, we choose to limit the sources in our sample.
\\

\begin{table}
    \centering
    \caption{Summary information of the compilation of quasars at high $z$ adopted in this work.}
    \label{tab:SMBH_data}
    \begin{tabular}{c|cccc}
         Name & $M (M_{\odot})$ & $z$ \\ 
        \hline
         J1342+0928 &  $(7.8^{+3.3}_{-1.9}) \times 10^8$ & 7.54  \\ 
         J1243+0100 & $(3.3 \pm 2) \times 10^8$          & 7.07 \\ 
         J1120+0641 & $(2.0^{+1.5}_{-0.7}) \times 10^9$    & 7.085 \\ 
         J0038-1527 & $(1.33 \pm 0.25)   \times 10^9$   & 7.021 \\
         J2348-3054 & $(2.1 \pm 0.5)  \times 10^9$   & 6.90 \\ 
         J0109-3047 & $(1.5 \pm 0.4)  \times 10^9$   & 6.80 \\
         J0305-3150 & $(9.5^{+0.8}_{-0.7})  \times 10^8$   & 6.61 \\
         P036+03    & $(1.9^{+1.1}_{-0.8})  \times 10^9$   & 6.54 \\
        \hline
    \end{tabular}
\end{table}

\textbf{Measurement of the Hubble parameter}: We adopt the latest measurements of the Hubble parameter obtained in a model-independent way. In particular, we use:
\begin{enumerate}

\item The re-analysis of the HST data using Cepheids as calibrators \citep{Riess_2019}, which led to a value $H_0 = 74.03 \pm 1.42 \, \mathrm{km \, s^{-1}\, Mpc^{-1}}$. We refer to this data point as R19. 
\item The recent determination of $H_0$ from the Tip-of-the-Red-Giant-Branch approach \citep{2019ApJ...882...34F}, which led to a value $69.6 \pm 2.0 \, \mathrm{km \, s^{-1}\, Mpc^{-1}}$, including systematic. We refer to this data point as F20.
\end{enumerate}

\textbf{BH growth model}: In our analysis we assume a flat-$\Lambda$CDM model as background scenario. Thus, our BH growth model is completely described by the following parameters: the seed initial mass $M_i$, the redshift formation of the BH $z_i$, the Hubble parameter $H_0$, the matter density parameter (baryons + dark matter) $\Omega_m$, the Eddington ratio $f_{\rm Edd}$ and the duty cycle ${\cal D}$.
To decrease the complexity of this 6-dimensional parameter space, we make the following, physically motivated assumptions: (i) we fix $\Omega_m=0.31$, (ii) we fix $z_i=25$, and (iii) we fix ${\cal D}=1$. The first assumption is motivated by the fact that the concordance flat-$\Lambda$CDM model seems to be characterized by smaller uncertainties on $\Omega_m$ than on $H_0$ \citep{2018arXiv180706209P}. The second assumption is supported by several models in early structure formation indicating that the formation of the first BHs occurred in the redshift range $20 \lesssim z \lesssim 30$ (e.g., \citealt{BL01, 2019arXiv191105791I}). As there is only a difference $\lesssim 100 \, \mathrm{Myr}$ in cosmic time within this redshift range, comparable to the typical quasar lifetime, we choose to fix $z_i=25$ as the mid-point of this range of interest. This is further supported by our preliminary analysis, during which we noticed that any value within the flat prior $z_i \in [20, 30]$ does not change our main results.
The third assumption stems from the fact that the Eddington ratio $f_{\rm Edd}$ and the duty cycle ${\cal D}$ are completely degenerate in any BH growth model. Hence, we assume ${\cal D}=1$ and interpret the Eddington ratio as averaged over sufficiently long timescales, typically of the order of the quasar lifetime, i.e. $\sim 100 \, \mathrm{Myr}$. This assumption is based on semi-analytical models (e.g., \citealt{Tanaka_2009}) as well as measurements of clustering of quasars at $z \gtrsim 6$ \citep{Shen_2007, Shankar_2009}.

We use the Markov Chain Monte Carlo (MCMC) method to analyze the parameters $\theta_i = \{ M_i, H_0, f_{\rm Edd}\}$, building  the posterior probability distribution function 

\begin{equation}
\label{L}
 p(D|\theta) \propto \exp \Big( - \frac{1}{2} \chi^2\Big) \, ,
\end{equation}
where 

\begin{equation}
\label{chi2_m}
\chi^2 = \sum_i^N \Big(\frac{M - M_{\rm th}}{\sigma_{M}} \Big)^2 + \Big(\frac{H_0 - H_{\rm 0,th}}{\sigma_{H_{0}}} \Big)^2 \,.
\end{equation}
Here, $N$ runs over all quasars in Table \ref{tab:SMBH_data}, $M_{\rm th}$, $M$ and $\sigma_M$ are the theoretical BH growth rate defined in Eq. (\ref{M}), the observational measurement and associated error on the mass of the quasars reported in Table \ref{tab:SMBH_data}, respectively. The quantities $H_0$ and $H_{\rm 0,th}$ represent the model-independent measurement of the Hubble parameter and its theoretical expectation inferred from the BH growth, respectively.

The goal of any MCMC approach is to draw $N$ samples $\theta_i$ from the general posterior probability density
\begin{equation}
\label{psd}
p(\theta_i, \alpha|D) = \frac{1}{Z} p(\theta,\alpha) p(D|\theta,\alpha)  \, ,
\end{equation}
where $p(\theta,\alpha)$ and $p(D|\theta,\alpha)$ are the prior distribution and the likelihood function, respectively. Here, the quantities $D$ and $\alpha$ are the set of observations and possible nuisance parameters. The amount $Z$ is a normalization term.

We subdivided our analysis in three steps: (i) we analyze the SMBH data only, assuming fixed values for the Eddington ratio $f_{\rm Edd}$; (ii) we analyze the SMBH data only with all parameters free; (iii) we consider the joint analysis SMBH + R19 and SMBH + F20 data. 

We perform the statistical analysis based on the \textit{emcee} algorithm \citep{2013PASP..125..306F}, assuming the theoretical model described in Sec. \ref{model} and the following priors on the parameters baseline: $H_0 \in [10, 90]$, $M_i \in [10^2, 10^5]$, and $f_{\rm Edd} \in [0.1, 1.5]$ in our overall analysis. We discard the first 20\% steps of the chain as burn-in. We follow the Gelman-Rubin convergence criterion \citep{R_test}, checking that all parameters in our chains have $R - 1 < 0.01$, where the parameter $R$ quantify the Gelman-Rubin statistic, also know as the potential scale reduction factor. It is recommended that $R < 1.1$ for all model parameters, in order to be confident that convergence is reached. We note $R < 1.01$ in our chains. We carry out a marginalization on $\Omega_m = 0.31$.

A flat prior on the seed mass $M_i \in [10^2, 10^5]$ allows both light and massive seeds (see our Sec. \ref{Introduction}) to be included in the analysis.
A flat prior on the Eddington ratio $f_{\rm Edd} \in [0.1, 1.5]$ allows us to consider both sub-Eddington accretion rates and super-Eddington accretion rates, up to $50\%$ above Eddington. Super-Eddington accretion rates are predicted to be common at high-$z$, due to a large availability of cold gas. Simple estimates presented in \cite{Begelman_Volonteri_2017} suggest that a fraction $\sim 10^{-3}$ of active galactic nuclei could be accreting at super-Eddington rates already at $z\sim 1$.

\section{Results and Discussion}
\label{results}
In the following we explore the parameter space $(M_i, H_0, f_{\rm Edd})$ with our MCMC approach, in order to constrain the probability distribution of the main parameters characterizing the growth of early BHs.

\begin{figure}
\centering
\includegraphics[scale=0.5]{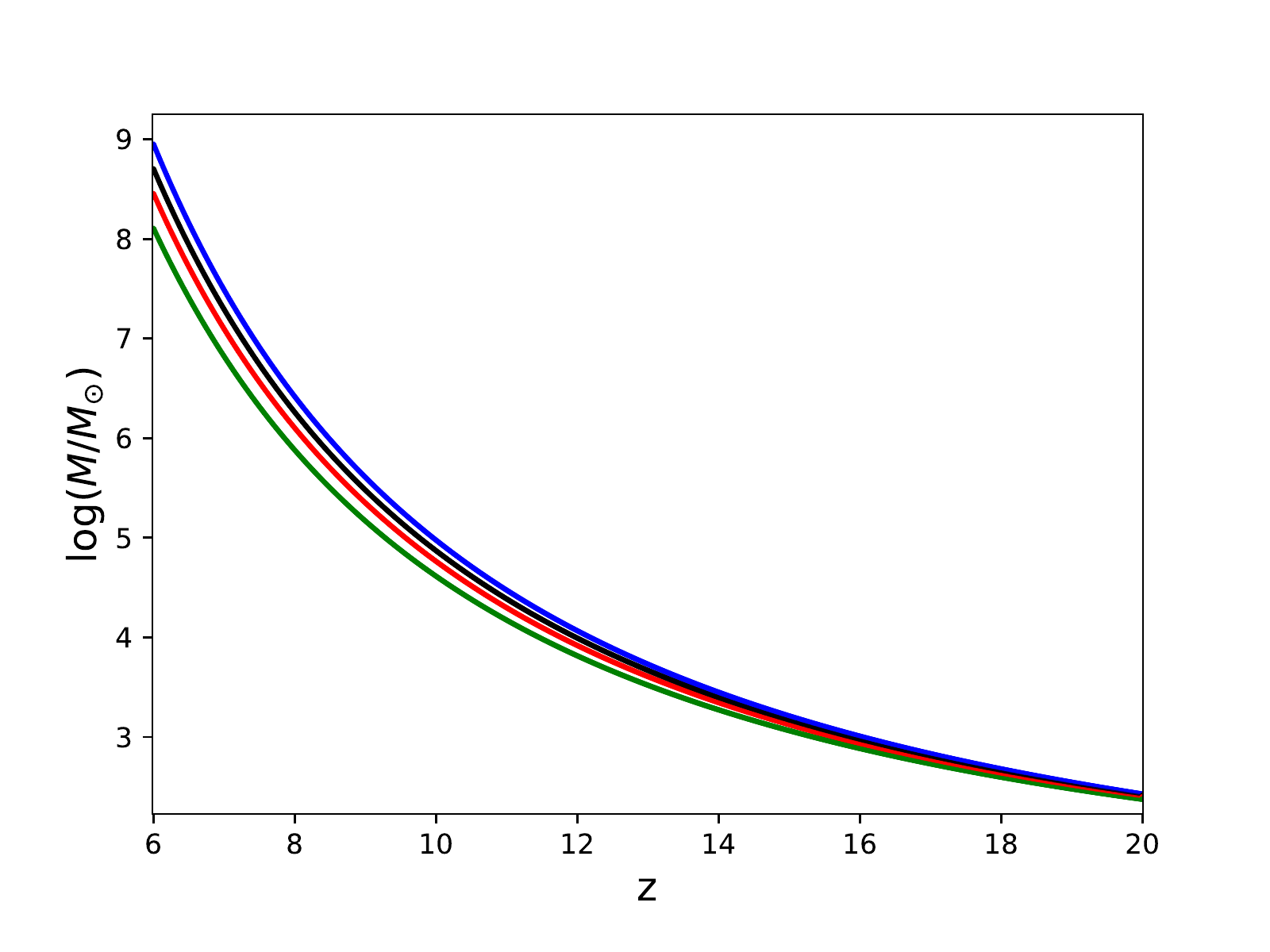} 
\caption{BH mass evolution as a function of redshift, assuming different values of the Hubble parameter: $H_0$ = 65 km s$^{-1}$Mpc$^{-1}$, $H_0$ = 67.4 km s$^{-1}$Mpc$^{-1}$ (Planck best fit value), $H_0$ = 70 km s$^{-1}$Mpc$^{-1}$ and $H_0$ = 74 km s$^{-1}$Mpc$^{-1}$ in blue, black, red and green, respectively. We assume as input values $M_i = 100 \Msun$ at $z_i = 25$, $\Omega_m = 0.31$, $f_{\rm Edd} = 1$ and ${\cal D} = 1$.}
\label{M1}
\end{figure}

\begin{figure}
\centering
\includegraphics[scale=0.5]{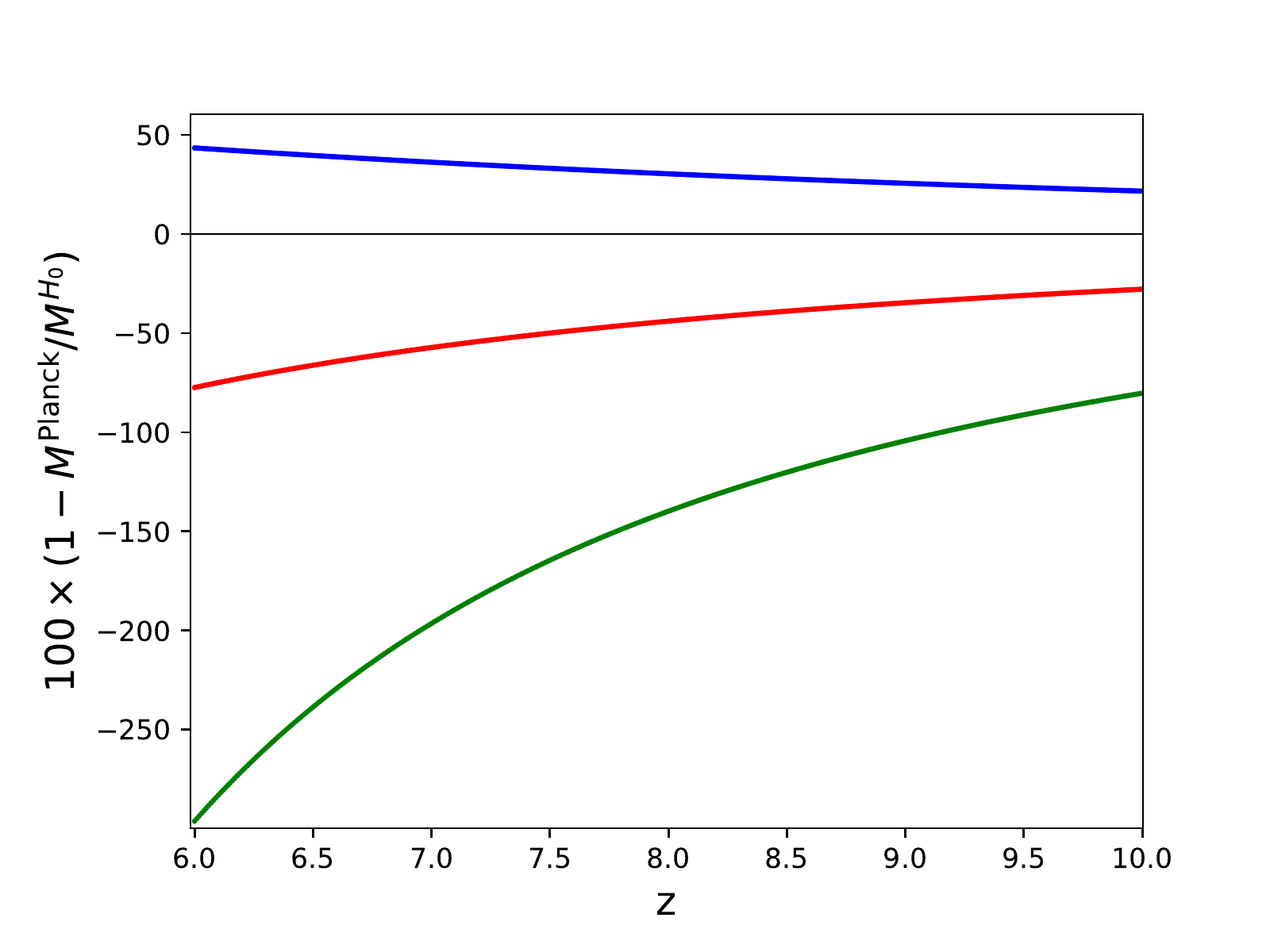} 
\caption{Percentage difference between $\Mblack^{\rm Planck}$ and $\Mblack^{H_0}$ as a function of redshift in the range $z \in$ [6, 10]. $\Mblack^{\rm Planck}$ represents the value of mass computed assuming the best-fit estimate of $H_0$ from Planck and $\Mblack^{H_0}$ the mass computed assuming $H_0=$ 65, 70, 74 km s$^{-1}$Mpc$^{-1}$, in blue, red and green, respectively. The black, thin line represents $\Mblack^{\rm Planck}$.}
\label{M2}
\end{figure}

As a case study, in Fig. \ref{M1} we show an example of BH mass growth as a function of $z$ for four values of $H_0$. For simplicity, we assume $f_{\rm Edd} = 1$, ${\cal D} = 1$ and initial seed mass of $M_i = 100 \Msun$ at $z_i=25$.
In Fig. \ref{M2} we show the percentage difference between $\Mblack^{\rm Planck}$ and $\Mblack^{H_0}$ (i.e., the mass computed assuming the Planck's value of $H_0$ or a generic value) as a function of redshift in the range $z \in$ [6, 10], in order to quantify how different $H_0$ values can influence the cosmic evolution of $\Mblack$. The differences in the BH mass evolution for $z \gtrsim 15$ are minimal, as the cosmic time from the seeding redshift $z_i \sim 25$ is very short, i.e. $\Delta t \sim 100 \, \mathrm{Myr}$.
On the contrary, we note that for $z \lesssim 15$ the differences in the predicted BH mass start to become significant. In particular, assuming different $H_0$ values and keeping $\Omega_m$ constant, we note significant changes in the final mass at a given $z$ value. Specifically, Fig. \ref{M2} shows that at $z=6$ the BH mass computed assuming $H_0=65$ km s$^{-1}$Mpc$^{-1}$ and $H_0=74$ km s$^{-1}$Mpc$^{-1}$ differs by $\sim$48\% and $\sim$300\%, respectively, from the mass computed assuming the fiducial Planck's value.
As all SMBHs observed thus far are at $z < 10$ and given the current tension in the measurement of $H_0$, analyzing how the value of this parameter can affect the BH mass evolution and, conversely, how observations of quasars can constrain the cosmological parameters is certainly very relevant.

\begin{figure}
\centering
\includegraphics[scale=0.5]{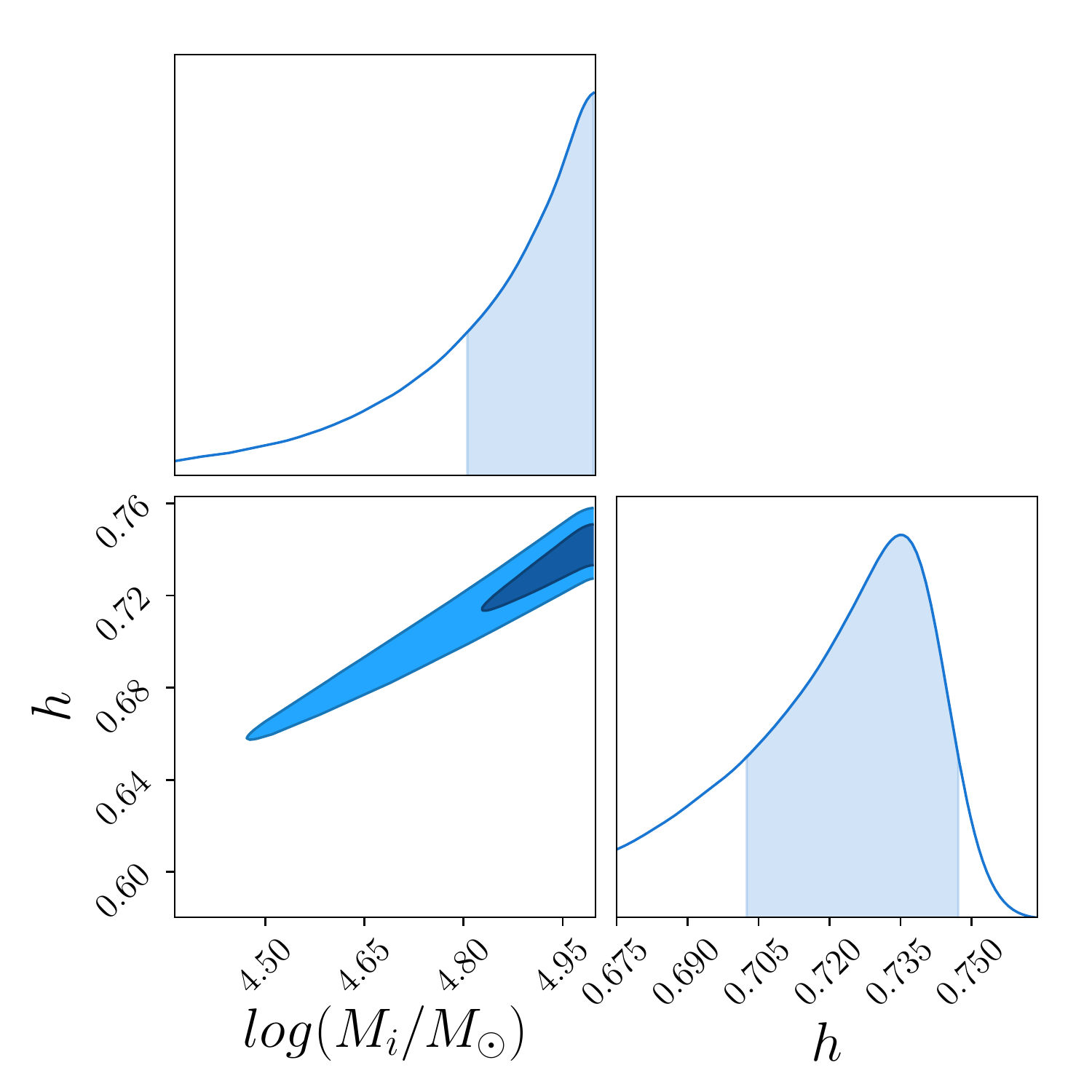} 
\caption{Two-dimensional, marginalized distributions in the parametric space $\log(M_{i}/M_{\odot}) - h$ at 1$\sigma$ and 2$\sigma$ CL from our SMBHs compilation data. We fix $f_{\rm Edd}=0.7$ and indicate with $h$ the reduced Hubble parameter, $H_0/100$, in units of km s$^{-1}$Mpc$^{-1}$.} 
\label{A1}
\end{figure}

As a first step in our statistical analysis, we only vary the seed mass $M_i$ and $H_0$, fixing the Eddington ratio at some physically motivated value, specifically at $f_{\rm Edd} = 0.7$. This sustained rate is supported by the fact that at $z \gtrsim 6$ most of the BHs are predicted to be accreting close to, or even above, the Eddington rate \citep{Begelman_Volonteri_2017}, due to a large availability of cold gas. Because of our assumption of ${\cal D} = 1$, we keep the value of $f_{\rm Edd}$ below unity.
In Fig. \ref{A1} we show the parametric space in the plane $\log(M_{i}/M_{\odot}) - h$, where $h$ is the reduced Hubble parameter. We find $\log(M_{i}/M_{\odot}) > 4.5$ at 95\% confidence level (CL).
This result clearly indicates a preference for heavy seeds over light seeds to match the observation of the earliest quasars.
For the Hubble parameter, we find $H_0 = 73.6^{+1.2}_{-3.3}$ km s$^{-1}$Mpc$^{-1}$ at 68\% CL, in substantial agreement with the measurement by \cite{Riess_2019}. Considering the 95\% CL bounds, instead, we note that $H_0$ may extend to lower values, which are compatible with high-$z$ measurements from the CMB.
As mentioned in Sec. \ref{Introduction}, there is a significant tension on $H_0$ when comparing very high-$z$ measures from CMB data with local measures. 

\begin{figure*}
\centering
\includegraphics[scale=0.38]{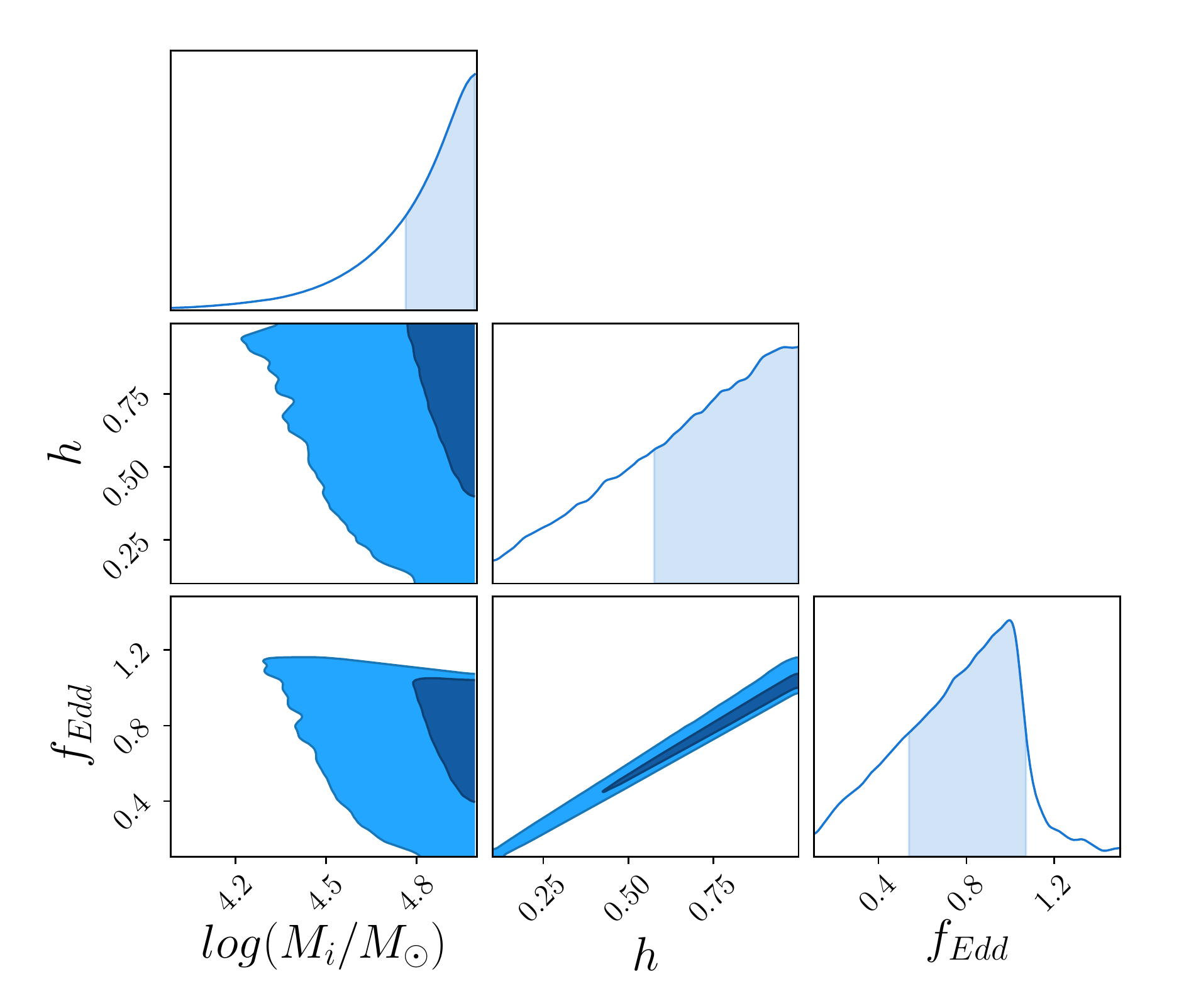} 
\includegraphics[scale=0.38]{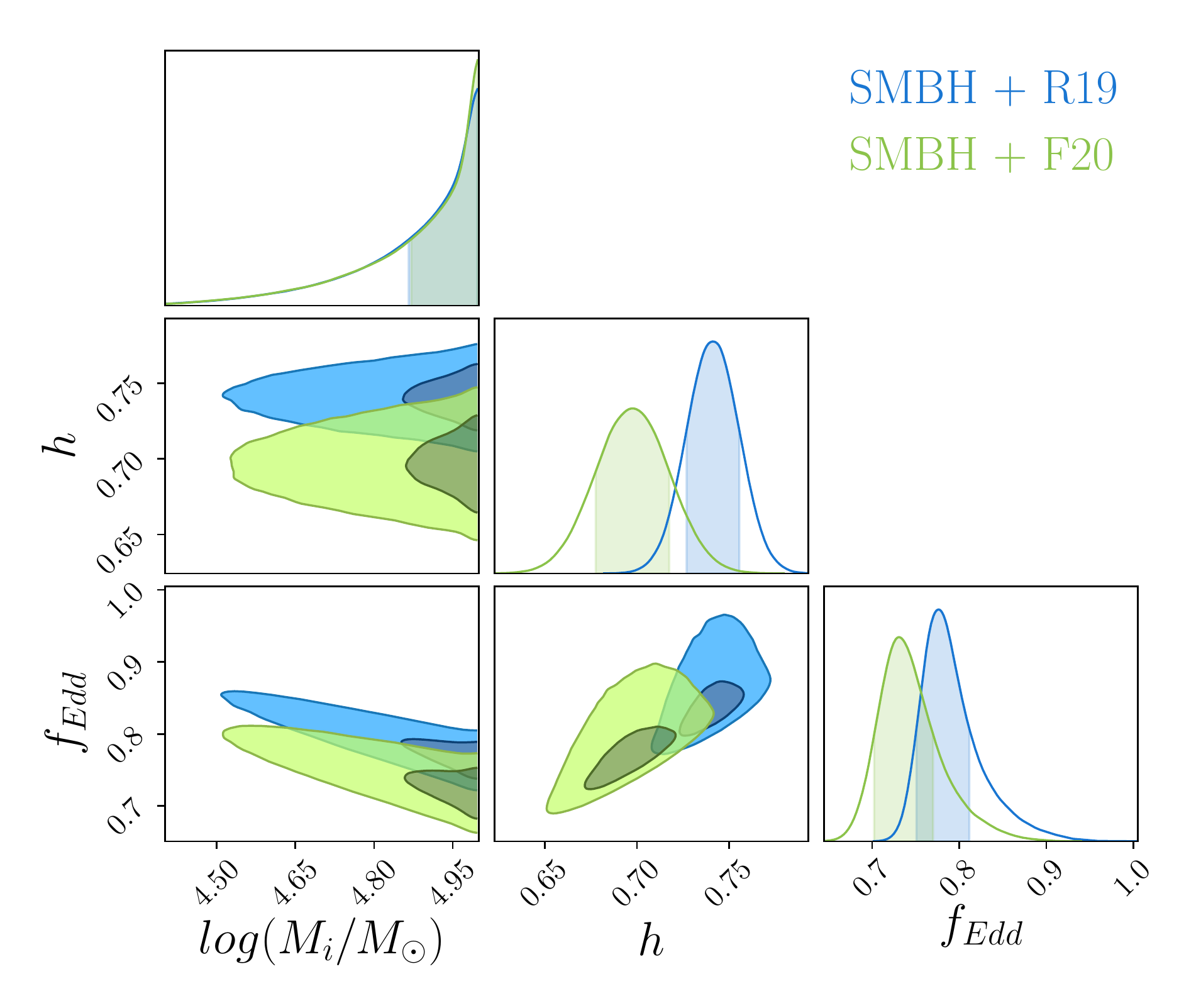} 
\caption{Left panel: Two-dimensional marginalized distributions of the free parameters $M_i$, $H_0$, $f_{\rm Edd}$ at 1$\sigma$ (darker area) and 2$\sigma$ CL (lighter area) from SMBHs data only. Here, $h$ is the reduced Hubble parameter, $H_0/100$ in units of km s$^{-1}$Mpc$^{-1}$. Right panel: Same as in left panel, but from the joint analysis SMBH + gaussian priors on $H_0$.}
\label{A2}
\end{figure*}

As a second step, we vary the parameters of the entire space $M_i$, $H_0$ and $f_{\rm Edd}$, while fitting the data for our sample of $z \gtrsim 6.5$ SMBHs. The results are shown in the left panel of Fig. \ref{A2}. In this global analysis, our results are as follows: $\log(M_{i}/M_{\odot}) > 4.26$, $H_0 > 55$ km s$^{-1}$Mpc$^{-1}$ and $f_{\rm Edd} = 0.995^{+0.076}_{-0.456}$. The lower bound on the initial mass is at 95\% CL, while the other constraints are at 68\% CL.
If we leave all three parameters of our BH growth model free, we notice that the constraints on $H_0$ are loose, certainly not competitive when compared to other robust cosmological tests. In principle, fixing certain parameters in the BH growth model can lead to robust constraints on $H_0$, when fitting to data from the farthest quasars. Unfortunately, constraining growth parameters is far from straightforward, as they can significantly vary from one BH to another.

In order to improve the estimates in our baseline parameters, we analyze the data combination SMBH + R19 and SMBH + F20. It is important to emphasize that R19 and F20 are model-independent measures. We are using these additional priors to break the statistical degeneracy which characterizes our analysis shown in the left panel of Fig. \ref{A2}, especially in the $H_0$ parameter, where we use a very loose and flat prior: $H_0 \in [10, 90]$. The R19 and F20 measures acts to keep our constraints for $H_0$ close to the observed values and, additionally, to improve the constraints on the other parameters. In fact, the BH growth parameters have significant positive correlations with $H_0$, mainly $f_{\rm Edd}$ when analyzed from SMBHs data only (see the left panel of Fig. \ref{A2}).

The right panel of Fig. \ref{A2} shows the parametric space from the joint analysis SMBHs + R19 and SMBH + F20. For the combination SMBH + R19, we find: $\log(M_{i}/M_{\odot}) > 4.5$ (at 95\% CL), $H_0 = 74^{+1.5}_{-1.4}$ km s$^{-1}$Mpc$^{-1}$ and $f_{\rm Edd}=0.777^{+0.035}_{-0.026}$  at 68\% CL. In the joint case SMBH + F20, we find: $\log(M_{i}/M_{\odot}) > 4.5$ (at 95\% CL), $H_0 =  69.7^{+2.0}_{-2.1}$ km s$^{-1}$Mpc$^{-1}$ and $f_{\rm Edd}=0.730^{+0.040}_{-0.028}$ at 68\% CL. We summarize our results for SMBH + R19 and SMBH + F20 as follows:

\begin{itemize}
    \item The addition of R19 and F20 data improves the constraints on $M_i$ by $\sim$ 0.3 dex. Our analysis show that the presence of quasars at $z \gtrsim 6.5$ strongly favors the formation of BH seeds with mass $\Mblack > 10^4 \Msun$.
    \item Regarding the parameters of the BH growth model, we note an improvement by 10.86\% and 9.58\% on $f_{\rm Edd}$ from the addition of R19 and F20 data, respectively.
\end{itemize}
This analysis confirms that Hubble parameter data can be fundamental to improve the constraints on the parameters of the BH growth model. 

As a complementary information to the best fit values, Table \ref{tab:parameter_correlations} reports the correlation matrix for the parameters resulting from the analysis SMBH + R19 (the corresponding table for SMBH + F20 is very similar). We emphasize that there is a strong anti-correlation between the parameters $M_i$ and $f_{\rm Edd}$, as expected from their physical interpretation. Also, we note a positive-correlation between $f_{\rm Edd}$ and $H_0$. This warrants a careful choice of the combination of these parameters whenever running numerical simulations.

\begin{table}
    \centering
    \caption{Correlation matrix for the parameters of the BH mass growth model (SMBH + R19 data).}
    \label{tab:parameter_correlations}
    \begin{tabular}{c|ccc}
         & $\log(M_{i}/M_{\odot})$ & $h$ & $f_{\rm Edd}$\\ 
        \hline
        $\log(M_{i}/M_{\odot})$ &  1.00 & -0.01 & -0.89 \\ 
                           $h$ & -0.01 &  1.00 &  0.43 \\ 
                     $f_{Edd}$ & -0.89 &  0.43 &  1.00 \\ 
        \hline
    \end{tabular}
\end{table}

\section{Final Remarks}
\label{conclusion}

We have investigated the effect that the value of the Hubble parameter $H_0$ has on models for the cosmic growth of BH seeds, by using an MCMC technique to fit mass measurements for $z \geq 6.5$ quasars. First, we noted that the predicted mass for a BH at $z=6$ changes by $> 300$\% if $H_0$ is changed from 65 to 74 $\mathrm{km \, s^{-1}\, Mpc^{-1}}$.  Assuming that seed formation occurs at $z\sim 25$, we find a strong preference for heavy seeds with $\log(M_{i}/M_{\odot}) > 4$ in all our models. With the specific priors on quasars used, light seed formation scenarios are rejected in our model at $\sim 3\sigma$. Our analysis is improved by considering gaussian priors on the value of $H_0$, specifically $H_0 = 74.03 \pm 1.42 \, \mathrm{km \, s^{-1}\, Mpc^{-1}}$ (\citealt{Riess_2019}, or R19) and $69.6 \pm 2.0 \, \mathrm{km \, s^{-1}\, Mpc^{-1}}$ (\citealt{2019ApJ...882...34F}, or F20).
When considering the joint analysis SMBH + R19 and SMBH + F20, the priors on the value of the Hubble parameter significantly improve the constraints on $f_{\rm Edd}$. We find that the Eddington ratio can be estimated with an accuracy of $\sim$3.9\% and $\sim$4.6\% from SMBH + R19 and SMBH + F20, respectively. Without targeted priors on $H_0$, we can constrain $f_{\rm Edd}$ only with an accuracy of $\sim$33\%. Of course, additional efforts are needed to better constrain the parameters of BH growth model: numerical and semi-analytical simulations, as well as additional data sets for high-$z$ quasars, will provide better estimates.
On the other hand, we showed that measurements of the Hubble parameter are fundamental to improve the constraints on these parameters. 

We recognize that there is a fundamental difference between the BH growth parameters and the value of the Hubble parameter. While the former parameters depend on the local accretion conditions and, in general, each BH is characterized by a different, possibly time-variable, combination of $(M_i, f_{\rm Edd})$, the latter is a constant value. Hence, constraining the Hubble parameter by improving our knowledge on $(M_i, f_{\rm Edd})$ seems realistic only with a very large statistical sample of high-$z$ quasars.
Despite this, we showed that there is a strong correlation between the Hubble parameter and the BH growth parameters: their combined analysis can thus bring a new perspective in the study of the farthest quasars.

An additional contribution to cosmology from the study of quasars could come from low-$z$ observations.
The Universe is currently undergoing an accelerated expansion and a complete explanation of this observation is still lacking. The concordance model thus far calls for an exotic component of dark energy, at low-$z$, to accelerate the cosmological expansion. Concurrently, the vast majority of AGNs (or accreting SMBHs at the center of galaxies) is currently detected at low redshift. Therefore, it could be fruitful to study how a background expansion of the Universe in the presence of well-motivated dark energy models can also influence SMBH mass estimates at low-$z$, and vice versa. As the lookback time is very sensitive to the density parameter of dark energy at low $z$, a modified background expansion could significantly change the mass growth of BHs in the nearby Universe. We defer this investigation to a future communication.

\section*{Acknowledgements}

\noindent
The authors thank the referee for her/his very constructive comments and suggestions. RCN would like to thank the Brazilian Agency FAPESP for financial  support under Project No. 2018/18036-5. FP acknowledges support from a Clay Fellowship administered by the Smithsonian Astrophysical Observatory and from the Black Hole Initiative at Harvard University, which is funded by grants from the John Templeton Foundation and the Gordon and Betty Moore Foundation.

%%----------------------------------------------------------------------------------------------------------------------
\bibliographystyle{mnras}

\label{lastpage}
\end{document}